\documentclass[a4paper,conference]{IEEEtran}
\IEEEoverridecommandlockouts
\usepackage{cite}
\usepackage{amsmath,amssymb,amsfonts}
\usepackage{algorithmic}
\usepackage{algorithm}
\usepackage{graphicx}
\usepackage{subfigure}
\usepackage{textcomp}
\usepackage{xcolor}
\usepackage{booktabs}
\usepackage{caption}
\usepackage{subcaption}
\def\BibTeX{{\rm B\kern-.05em{\sc i\kern-.025em b}\kern-.08em
    T\kern-.1667em\lower.7ex\hbox{E}\kern-.125emX}}
\makeatletter
\renewcommand{\maketag@@@}[1]{\hbox{\m@th\normalsize\normalfont#1}}
\makeatother
\begin{document}
\title{End-to-End Generative Semantic Communication Powered by Shared Semantic Knowledge Base}
\author{
	\IEEEauthorblockN{
		Shuling Li\IEEEauthorrefmark{1}, 
		Yaping Sun\IEEEauthorrefmark{2}\IEEEauthorrefmark{3}, 
		Jinbei Zhang\IEEEauthorrefmark{1}, 
		Kechao Cai\IEEEauthorrefmark{1},
		Shuguang Cui\IEEEauthorrefmark{3}\IEEEauthorrefmark{2}
        and Xiaodong Xu\IEEEauthorrefmark{4}\IEEEauthorrefmark{2}}
        	\IEEEauthorblockA{\IEEEauthorrefmark{1}Sun Yat-sen University, Shenzhen, China}
	\IEEEauthorblockA{\IEEEauthorrefmark{2}Department of Broadband Communication, Peng Cheng Laboratory, Shenzhen, China}
	\IEEEauthorblockA{\IEEEauthorrefmark{3}Future Network of Intelligent Institute, Chinese University of Hong Kong (Shenzhen), Shenzhen, China} 
	\IEEEauthorblockA{\IEEEauthorrefmark{4}Beijing University of Posts and Telecommunications, Beijing, China}
        \IEEEauthorblockA{Emails: lishling8@mail2.sysu.edu.cn, sunyp@pcl.ac.cn, \{zhjinbei, caikch3\}@mail.sysu.edu.cn,\\  shuguangcui@cuhk.edu.cn, xuxiaodong@bupt.edu.cn}
        }
\maketitle
\begin{abstract}
Semantic communication has drawn substantial attention as a promising paradigm to achieve effective and intelligent communications.
However, efficient image semantic communication encounters challenges with a lower testing compression ratio (CR) compared to the training phase. 
To tackle this issue, we propose an innovative semantic knowledge base (SKB)-enabled generative semantic communication system for image classification and image generation tasks. 
Specifically, a lightweight SKB, comprising class-level information, is exploited to guide the semantic communication process, which enables us to transmit only the relevant indices. This approach promotes the completion of the image classification task at the source end and significantly reduces the transmission load.
Meanwhile, the category-level knowledge in the SKB facilitates the image generation task by allowing controllable generation, making it possible to generate favorable images in resource-constrained scenarios. Additionally, semantic accuracy is introduced as a new metric to validate the performance of semantic transmission powered by the SKB.
Evaluation results indicate that the proposed method outperforms the benchmarks and achieves superior performance with minimal transmission overhead, especially in the low SNR regime.

\end{abstract}
\begin{IEEEkeywords}
Generative semantic communication, semantic knowledge base (SKB), image transmission, image generation, deep learning.
\end{IEEEkeywords}

\section{Introduction}
As a new architecture where both application requirements and information are considered in transmission process, semantic communication is a promising paradigm to empower “connected intelligence” in 6G\cite{letaiefEdgeArtificialIntelligence2022, yangSemanticCommunicationsFuture2022}. Focused on conveying task-relevant information rather than syntactic information, semantic communication can significantly reduce transmission overhead and improve reliability compared to conventional communication\cite{yangSemanticCommunicationsFuture2022}. Specifically, with the support of \textit{semantic knowledge base (SKB)}, the task-irrelevant information in source message is removed and only the crucial meaning about the specific task (\textit{semantic information}) is transmitted. 
However, under some unfavorable conditions, such as lower testing CR than that in the training phase, the semantic transmission is hindered. 
We attempt to cope with this challenge via an SKB-enabled generative semantic communication system, which can provide promising outputs in scenarios with limited transmission overhead.

There have been some studies considering the construction of SKB, mainly for image data and text data. For text transmission, some works define SKB as some brief information extracted from the text data, i.e., knowledge graph\cite{zhouCognitiveSemanticCommunication,xuofdm} and sentences carefully selected from the  dataset\cite{yiSemanticCommunicationSystems2023}. 
Similarly, for image transmission, there are works have defined SKB as high-level representations from the training data \cite{zhangWynerZivCodingbasedSemantic,huRobustSemanticCommunications,sunZeroshotMultilevelFeature}. 
With the similar thinking of building SKB in \cite{yiSemanticCommunicationSystems2023}, a new measure named Weighted Data-Semantic (WDS) distance is proposed for discovering the image quantization centroids in \cite{zhangWynerZivCodingbasedSemantic}. These quantization centroids are of the same size as the image data and form the SKB. 
In \cite{yiSemanticCommunicationSystems2023} and \cite{zhangWynerZivCodingbasedSemantic}, the source message and corresponding knowledge in the SKB are jointly encoded at the transmitter, and then only residual information is transmitted, which reduces the transmission consumption. However, the aforementioned works \cite{zhouCognitiveSemanticCommunication,xuofdm,yiSemanticCommunicationSystems2023,zhangWynerZivCodingbasedSemantic} establish a strong positive correlation between the size of SKB and that of the dataset, thus the SKB will occupy considerable storage space when the dataset is large. 
Sun \textit{et al.} \cite{sunZeroshotMultilevelFeature} utilize a succinct SKB composed of semantic knowledge vectors corresponding to image classes, and establish multi-level feature extractor, which supports multi-level feature transmission to execute remote zero-shot recognition task effectively.

Although the existing works leverage SKB to save transmission cost, most of them consider only one task at the receiver, leading to insufficient semantic utilization. In order to enhance the competence of the overall system, it is a preferable choice to explore multi-task semantic communication that can handle multiple tasks, such as image classification and image reconstruction tasks \cite{zhangDeepLearningEnabledSemantic2023, tangContrastiveLearningBased2023}.
The general process is reconstructing the image first and then classifying the recovered image. However, as the task-related information is transmitted, it is more effective to perform both tasks straightforwardly with semantic information. Moreover, most works recognize the metrics of tasks as whole system performance while ignore the metric of semantics, which cannot perfectly represent the semantic transmission efficiency.

In addition, for certain scenarios with low transmission load, how to enhance task performance with limited resources is a crucial challenge. 
In this case, the goal of communication is providing a satisfactory quality of experience (QoE) instead of perfect restoration of bits or pixels. Generative artificial intelligence (GAI), with its immense potential in digital generation \cite{ThePowerofGenerativeAI}, is expected to achieve this objective by generating an approximate distribution with source data as the transmission results \cite{Generativejscc}. Thus, embedding generative models into semantic communication emerges as a viable approach for performance improvement in adverse situations \cite{gaisem}.

\begin{figure*}[ht]
	\centering
	\includegraphics[width=0.8\linewidth]{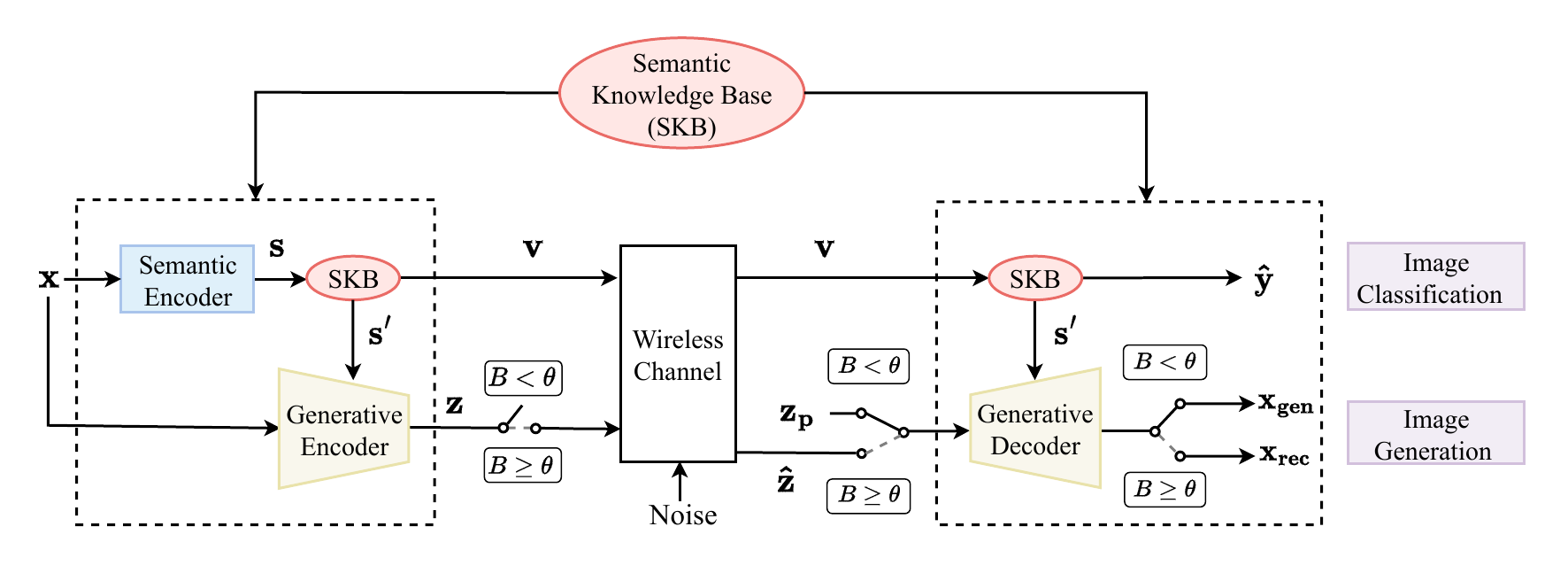}
	\caption{The framework of proposed SKB-enabled generative semantic communication system. All ellipses indicate the semantic knowledge base.} 
	\label{1-SKB_model}
\end{figure*}
Therefore, equipping SKB and generative models in semantic communication is of great significance to deal with challenging environments. 
In this paper, an SKB-enabled generative semantic communication system is proposed for image classification task and image generation task. 
Specifically, a concise SKB, representing the current understanding of the image data by both communication parties, is exploited to guide the communication process. By integrating conditional variational autoencoder (CVAE) into both sides, we acquire favorable results with extreme low overhead. The main contributions of this paper are summarized as follows.
\begin{itemize}
    \item An SKB-enabled semantic communication system is proposed, in which a powerful SKB instructs the transmitter to extract semantic feature and leads the receiver to execute intelligent tasks. The shared SKB enables the transmission of only the pertinent indices and encourages the execution of image classification task at the sender, thereby  greatly reducing the transmission load.
    \item Based on the aforementioned model, a CVAE-based generative semantic communication framework is further designed, capable of performing two typical tasks, i.e., image classification and image generation. 
    Besides, a new metric named semantic accuracy is introduced to measure the valuable information obtained from incorrect inference results.
    \item Numerical results demonstrate that the proposed system has superior performance on the two tasks with low transmission load and exhibits robustness against channel noise, as compared with benchmark schemes without guidance of SKB.  
\end{itemize}
\addtolength{\topmargin}{0.065cm}
\section{SKB-Enabled Generative Semantic Communication System}
In this section, the construction of SKB and the SKB-enabled semantic communication process are illustrated. As shown in Fig.~\ref{1-SKB_model}, the system model consists of two branches to perform image classification and image generation tasks, respectively.
Note that the image generation task in this paper refers to generating images with the same class of source or recovering the source images. Choice of the two modes depends on the testing CR $B$ and training CR $\theta$. In particular, if the testing CR $B<\theta$, only associated indices are transmitted to generate images belonging to the same class as the source images. Otherwise, if $B\geq\theta$, more detailed information is sent for image reconstruction at the receiver.

\subsection{Semantic knowledge base (SKB)}
As illustrated in Fig.~\ref{1-SKB_model}, the transmitter and receiver are endowed with a shared SKB. The SKB indicates some background knowledge or prior information, describing the implicit relations between the task and the data. Considering above elements and inspired by \cite{sunZeroshotMultilevelFeature}, we construct the SKB with class-level attribute vectors. Each attribute vector corresponds to a unique category and shows the states of certain attributes. Besides, the SKB at the transmitter is shared and synchronized with that at the receiver. In this way, the SKB reflects the current comprehension of both sides about the images.
Let $\mathcal{M}\overset{\Delta}{=}\{1,2,\cdots,M\}$ denote the set of classes, $d$ denote the number of attributes and $\mathbf{k}_{m} \in\mathbb{R}^{d}$ denote the attribute vector of the $m$-th class, respectively. Then, let $\mathcal{K}\overset{\Delta}{=}\{\mathbf{k}_{m}\in\mathbb{R}^{d} \}_{m\in \mathcal{M}}$ denote the set of attribute vectors, i.e., the SKB. 

\subsection{Transmitter}
Let $\mathbf{x}\in \mathbb{R}^{W\times H\times C}$ denote the input image, where $W$, $H$, $C$ denote the width, height and channel of the image, respectively. Under the supervision of the SKB, the semantic encoder $S_\alpha(\cdot)$ learns to extract pivotal information related to the category of $\mathbf{x}$, yielding semantic feature $\mathbf{s}\in \mathbb{R}^{d}$ with the same dimension as the attribute vectors in the SKB. The semantic feature $\mathbf{s}$ is intuitively regarded as the predicted attribute values of the input image. Then, these attribute information will be leveraged for both image classification and image generation tasks. 

Specifically, the distance between the semantic feature $\mathbf{s}$ and each attribute vector $\mathbf{k}_{m}$ in the SKB is measured
 by cosine similarity, and the vector with the maximum similarity is selected out as the most similar vector $\mathbf{s{'}}\in \mathbb{R}^{d}$, i.e.,
 \begin{equation}
     \mathbf{s'}=\arg\max_{\mathbf{k}_{m}\in\mathcal{K}}D(\mathbf{s},\mathbf{k}_{m}),
 \end{equation}
where $D(\mathbf{s},\mathbf{k}_{m})$ represents the cosine similarity between $\mathbf{s}$ and $\mathbf{k}_{m}$, i.e., $D(\mathbf{s},\mathbf{k}_{m})=\frac{<\mathbf{s},\mathbf{k}_{m}>}{ | \mid \mathbf{s}\mid\mid\times\mid\mid\mathbf{k}_{m}\mid\mid}$. $<\cdot,\cdot>$ denotes the inner product of two vectors. Then, the vector $\mathbf{s{'}}$ is what we intend to transmit, which is the semantic information corrected by the SKB. 
Since the SKB is shared between both ends, instead of transmitting the selected vector $\mathbf{s{'}}$, its corresponding index $\mathbf{v}$ in the SKB is chosen to be transmitted, which requires fewer bits. Let $t$ denote the number of symbols occupied by index v. In this paper, the index v is represented by at most 8 bits so $t = 1$. Given the small transmission overhead of the index, we assume that the transmission process of the index is error-free, similar to \cite{yiSemanticCommunicationSystems2023}. Besides, we assume the index of the attribute vector as the class label. Thus, the image classification task is accomplished at the transmitter.

For the image generation task, a CVAE is deployed to obtain the specific information of the images in the latent space, which consists of generative encoder $E_\delta(\cdot,\cdot)$ at the transmitter and generative decoder $E_\eta^{-1}(\cdot,\cdot)$ at the receiver. The input of the generative encoder $E_\delta(\cdot,\cdot)$ contains the source image $\mathbf{x}$ and the selected attribute vector $\mathbf{s{'}}$, where $\mathbf{s{'}}$ is the conditional input to make the generation process controllable. The output of the generative encoder $E_\delta(\cdot,\cdot)$ is denoted as $\mathbf{z}\in \mathbb{R}^{l}$, where $l$ denotes the dimension of $\mathbf{z}$.

We define CR as the ratio between the number of transmitted symbols and input symbols, which can be represented as $\frac{t+l}{W\times H\times C}$. If the testing CR $B<\theta$, only the index $\mathbf{v}$ will be sent to generate class-consistent images. Otherwise, $\mathbf{z}$ will be transmitted additionally to reconstruct $\mathbf{x}$.

\addtolength{\topmargin}{-0.13cm}
\subsection{Physical channel }
In this paper, additive white Gaussian noise (AWGN) channel is considered as the physical channel. 
Firstly, the index $\mathbf{v}$ is transmitted via an error-free link. Then, if $\mathbf{z}$ is additionally transmitted, the received signal $\mathbf{\hat{z}}\in \mathbb{R}^{l}$ is denoted as $\mathbf{\hat{z}}=\mathbf{z}+\mathbf{n}$, where $\mathbf{n}$ indicates the Gaussian noise, i.e., $\mathbf{n}\sim\mathcal{N}(0,\sigma^2\mathbf{I})$. 

\subsection{Receiver}
At the receiver, the predicted class $\hat{\mathbf{y}}$ is equal to the index $\mathbf{v}$ in terms of the numerical value. The semantic information $\mathbf{s'}$ is acquired by referring to the SKB with the index $\mathbf{v}$. 
For the image generation task, when $B<\theta$, the generative decoder $E_\eta^{-1}(\cdot,\cdot)$ utilizes $\mathbf{s'}$ and random Gaussian samples $\mathbf{z_{p}}\in \mathbb{R}^{l}$ to generate the image $\mathbf{x_{gen}}$, which is of the same category as $\mathbf{x}$. When $B\geq\theta$, by receiving some latent information via the channel as $\mathbf{\hat{z}}$ and decoding it with $\mathbf{s'}$, the original image $\mathbf{x}$ is reconstructed as $\mathbf{x_{rec}}$.
The end-to-end semantic communication process is given in Algorithm~\ref{alg:1}, where $\mathbf{x}_{i}$ denotes the $i$-th image in a batch and ${\mathbf{s}_{i}'}$ denotes the corresponding estimated semantic knowledge indexed by $\mathbf{v}_{i}$.

\begin{algorithm}[t]
	\renewcommand{\algorithmicrequire}{\textbf{Input:}}
	\renewcommand{\algorithmicensure}{\textbf{Output:}}
	\caption{The end-to-end semantic communication process}
	\label{alg:1}
    \begin{algorithmic}[1]
		\REQUIRE Image data $\mathcal{X}=\{\mathbf{x}_{i}\}$,  SKB $\mathcal{K}=\{\mathbf{k}_{m}\}_{{m}\in \mathcal{M}} $, testing CR $B$
		\ENSURE Predicted class $\hat{\mathbf{y}}_{i}$, generated image $\mathbf{x}_{i,\mathbf{gen}}$ or reconstructed image $\mathbf{x}_{i,\mathbf{rec}}$ 
		\FOR  {$\mathbf{x}_{i} \text{ in } \mathcal{X}$}
            \STATE $\mathbf{s}_{i} \leftarrow S_\alpha(\mathbf{x}_{i})$
            \STATE $\mathbf{v}_{i} \leftarrow $ Locate the attribute vector in the SKB with the greatest cosine similarity and return its index
            \IF{$B<\theta$} 
                \STATE Transmit $\mathbf{v}_{i}$ over the channel
            \ELSE 
                \STATE ${\mathbf{s}_{i}'} \leftarrow $ Retrieve the attribute vector in the SKB by the index $\mathbf{v}_{i}$
                \STATE $\mathbf{z}_{i} \leftarrow E_\delta(\mathbf{x}_{i},{\mathbf{s}_{i}'})$
                \STATE Transmit $\mathbf{v}_{i}$ and $\mathbf{z}_{i}$ over the channel
            \ENDIF 
            \STATE $\hat{\mathbf{y}}_{i} \leftarrow$ $\mathbf{v}_{i}$
            \STATE ${\mathbf{s}_{i}'} \leftarrow$ Retrieve the attribute vector in the SKB by $\mathbf{v}_{i}$
            \IF{$B<\theta$} 
                \STATE $\mathbf{z_p} \leftarrow $ Sample from Normal distributions 
                \STATE $\mathbf{x}_{i,\mathbf{gen}} \leftarrow E_\eta^{-1}(\mathbf{z_p},{\mathbf{s}_{i}'})$
            \ELSE     
                \STATE $\mathbf{x}_{i,\mathbf{rec}} \leftarrow E_\eta^{-1}(\hat{\mathbf{z}}_{i},{\mathbf{s}_{i}'})$
            \ENDIF 
		\ENDFOR
    \end{algorithmic}  
\end{algorithm}

\newcounter{TempEqCnt} 
\setcounter{TempEqCnt}{\value{equation}} 
\setcounter{equation}{2} 
\begin{figure*}[ht] 
	\begin{equation}\label{nvae} 
		\mathcal{L}_{\mathrm{VAE}}(\mathbf{x})=-\mathbb{E}_{q(\mathbf{z}|\mathbf{x})}\left[\log p(\mathbf{x}|\mathbf{z})\right]+
 \mathrm{KL}(q(\mathbf{z}_1|\mathbf{x})||p(\mathbf{z}_1))+\sum_{{g}=2}^{G}\mathbb{E}_{{q}(\mathbf{z}_{<g}|\mathbf{x})}\left[\mathrm{KL}(q(\mathbf{z}_g|\mathbf{x},\mathbf{z}_{<g})||p(\mathbf{z}_g|\mathbf{z}_{<g}))\right]
	\end{equation}
\begin{small} 
        \begin{equation}\label{cvae} 
             \begin{aligned}
	             \mathcal{L}_{\mathrm{2}}(\mathbf{x},\mathbf{k},\mathbf{z};\delta,\eta)=-\mathbb{E}_{q(\mathbf{z}|\mathbf{x,k})}\left[\log p(\mathbf{x}|\mathbf{z,k})\right]+\beta
 \mathrm{KL}(q(\mathbf{z}_1|\mathbf{x,k})||p(\mathbf{z}_1|\mathbf{k}))+\beta\sum_{\boldsymbol{g}=2}^{G}\mathbb{E}_{{q}(\mathbf{z}_{<g}|\mathbf{x,k})}\left[\mathrm{KL}(q(\mathbf{z}_g|\mathbf{x},\mathbf{z}_{<g},\mathbf{k})||p(\mathbf{z}_g|\mathbf{z}_{<g},\mathbf{k}))\right] 
             \end{aligned}
	\end{equation}
 \end{small}
 \hrulefill  
\end{figure*}

\section{Model Implementation}
In this section, the implementation details of the system model are introduced, including both loss function and training strategy. Besides, the performance of semantic transmission is considered by a metric named semantic accuracy.

\subsection{Loss function and training strategy}
The objective of the proposed system is to accurately restore the transmitted information at semantic level and accomplish tasks under a given transmitted overhead. That is, obtaining representative and well-organized semantic information and ensuring its reliable transmission. Specifically, since the semantic encoder learns to extract semantic feature with the instruction of the SKB, the difference between the semantic feature $\mathbf{s}$ and corresponding attribute vector $\mathbf{k}_{m}$ needs to be minimized. Motivated by the attribute prototype network (APN) \cite{xuAttributePrototypeNetwork} which shows remarkable proficiency in recognizing image attributes, the loss function of semantic encoder $S_\alpha(\cdot)$ is designed as
\newcounter{TempEqCnt1} 
\setcounter{TempEqCnt1}{\value{equation}} 
\setcounter{equation}{1} 
\begin{equation}
    \begin{aligned}
	\mathcal{L}_{{1}}(\mathbf{s},\mathbf{k};\alpha)= 
	\lambda_1\frac1N&\sum_{i=1}^N(\mathbf{s}_{i}-\mathbf{k}_{m_i})^2\\ &+\lambda_2\log\frac{\exp<\mathbf{s}_{i},\mathbf{k}_{m_i}>}{\sum\limits_{i=1}^N \exp<\mathbf{s}_{i},\mathbf{k}_{m_i}>},
    \end{aligned}
\end{equation}
where $\mathbf{s}_{i}$ denotes the semantic feature of $i$-th image in a batch of $N$ images, $m_i$ denotes the corresponding class index of that image, and $\mathbf{k}_{m_i}$ denotes the attribute vector of the $m_i$-th class in the SKB. $\lambda_1$ and $\lambda_2$ denote the weights of the two terms respectively. The CNN-based prototype network \cite{xuAttributePrototypeNetwork} is served as the semantic encoder in this paper. In addition, the lossless transmission of the relevant indices guarantees the reliable transfer of semantic content. 

For the generative codec, the semantic consistency between the input and output can be achieved by aligning their distributions to be as similar as possible. VAE is a generative model that aims to minimize the discrepancy between the joint distributions from encoder and that from decoder, which coincides with our purpose. We adopt a deep hierarchical VAE as the framework, referencing Nouveau VAE (NVAE) \cite{NEURIPS2020_e3b21256}, which exhibits outstanding image generation capability with multi-scale coding structure. 
Specifically, the latent variables are split into $G$ separate groups, i.e., $\mathbf{z}=\{\mathbf{z}_1,\mathbf{z}_2,...,\mathbf{z}_G\}$. Thus, the prior is denoted as $p(\mathbf{z})=\prod_{g}p(\mathbf{z}_g|\mathbf{z}_{<g})$ and the approximate posterior as $q(\mathbf{z}|\mathbf{x})=\prod_gq(\mathbf{z}_g|\mathbf{z}_{<g},\mathbf{x})$, where each conditional probability in both formulas is represented as factorial Normal distribution. 
Assume the $j$-th variable in $\mathbf{z}_{g}$ is a Normal distribution, i.e., $p(z_g^j|\mathbf{z}_{<g})=\mathcal{N}\left(\mu_j(\mathbf{z}_{<g}),\sigma_j(\mathbf{z}_{<g})\right)$. The relative approximate posterior is designed as $q(z_g^j|\mathbf{z}_{<g},\mathbf{x})=\mathcal{N}\left(\mu_j(\mathbf{z}_{<g})+\Delta\mu_j(\mathbf{z}_{<g},\mathbf{x}),\sigma_j(\mathbf{z}_{<g})\cdot\Delta\sigma_j(\mathbf{z}_{<g},\mathbf{x})\right)$, where $\Delta\mu_j(\mathbf{z}_{<g},\mathbf{x})$ and $\Delta\sigma_j(\mathbf{z}_{<g},\mathbf{x})$ are computed by neural networks. Then, the loss function of the hierarchical VAE is formulated as (\ref{nvae}), consisting of reconstruction loss and regularization loss. 
In order to make the generation content controllable, we extend the hierarchical VAE to hierarchical CVAE, in which the attribute information is inputted as condition. Therefore, the loss function of our generative model $E_\delta(\cdot,\cdot)$ and $E_\eta^{-1}(\cdot,\cdot)$ is formulated as (\ref{cvae}), where $\beta$ is a hyper-parameter.

We train the above two modules separately. The detailed training processes are shown in Algorithm~\ref{alg:2} and Algorithm~\ref{alg:3}, where $\mathbf{x}_{i,{m_i}}$ denotes the $i$-th image in a batch and $m_i$ denotes its class index.
$\alpha$, $\delta$ and $\eta$ are trainable parameters. The semantic encoder learns to extract class-level attributes from images via training. The generative encoder and  decoder are jointly trained under the supervision of the attribute vectors in the SKB. Thus, the generative decoder acquires the ability to generate controlled images.

\begin{algorithm}[t]
	\renewcommand{\algorithmicrequire}{\textbf{Input:}}
	\renewcommand{\algorithmicensure}{\textbf{Output:}}
	\caption{Training of the semantic encoder $S_\alpha(\cdot)$}
	\label{alg:2}
    
    \begin{algorithmic}[1]
		\REQUIRE Training data $\mathcal{X}=\{{\mathbf{x}_{i,{m_i}}}\}$, SKB $\mathcal{K}=\{\mathbf{k}_{m_i}\}_{{m_i}\in \mathcal{M}} $
		\ENSURE $S_\alpha(\cdot)$
		\FOR  {$\mathbf{x}_{i,{m_i}} \text{ in } \mathcal{X}$}
            \STATE $\mathbf{s}_{i} \leftarrow S_\alpha(\mathbf{x}_{i,{m_i}})$
            \STATE Compute $\mathcal{L}_{\text{1}}$ by (2) 
            \STATE Update $\alpha$ via Adam
        \ENDFOR
    \end{algorithmic}  
\end{algorithm}

\begin{algorithm}[t]
	\renewcommand{\algorithmicrequire}{\textbf{Input:}}
	\renewcommand{\algorithmicensure}{\textbf{Output:}}
	\caption{Training of the generative encoder $E_\delta(\cdot,\cdot)$ and generative decoder $E_\eta^{-1}(\cdot,\cdot)$ }
	\label{alg:3}
    
   \begin{algorithmic} [1]
        \REQUIRE Training data $\mathcal{X}=\{{\mathbf{x}_{i,{m_i}}}\}$, SKB $\mathcal{K}=\{\mathbf{k}_{m_i}\}_{m_i\in \mathcal{M}}$
        \ENSURE $E_\delta(\cdot,\cdot)$, $E_\eta^{-1}(\cdot,\cdot)$
        \FOR {$\mathbf{x}_{i,{m_i}} \text{ in } \mathcal{X}$}
            \STATE $\mathbf{k}_{m_i} \leftarrow $ Find the corresponding attribute vector in the SKB
            \STATE $\mathbf{z}_i \leftarrow E_\delta(\mathbf{x}_{i,m_i},\mathbf{k}_{m_i})$
            \STATE Transmit $\mathbf{z}$ over the channel
            \STATE $\mathbf{x}_{i,\mathbf{rec}} \leftarrow E_\eta^{-1}(\hat{\mathbf{z}}_i,\mathbf{k}_{m_i})$
            \STATE Compute $\mathcal{L}_{\text{2}}$ by (4) 
            \STATE Update $\delta$, $\eta$ via Adam
		\ENDFOR
    \end{algorithmic}  
\end{algorithm}

\subsection{Performance metric}
In order to measure the effectiveness of semantic transmission, we consider semantic accuracy as a new performance metric, which indicates the accuracy of the received semantic information. Specifically, given an input image $\mathbf{x}$ with the $m$-th class, by the comparison between the received semantic information $\mathbf{s'}=[s'_{1},s'_{2},\cdots,s'_{d}]$ and the real attribute feature $\mathbf{k}_{m}=[k_{m_1},k_{m_2},\cdots,k_{m_d}]$, semantic accuracy is calculated as the proportion of correctly predicted attributes, i.e., 
\newcounter{TempEqCnt3} 
\setcounter{TempEqCnt3}{\value{equation}} 
\setcounter{equation}{4} 
\begin{equation}
  Semantic\ accuracy = \frac{\sum\limits_{{i}=1}^{d} [|{s'_{i}}-{k_{m_i}}| \leq \gamma]}d,
\end{equation}
where $s'_{i}\in[0,1]$, ${k_{m_i}}\in[0,1]$ and $\gamma$ is set as 0.0005. $[|{s'_{i}}-{k_{m_i}}| \leq \gamma]$ denotes an indicator function that equals 1 if $|{s'_{i}}-{k_{m_i}}| \leq \gamma$, and 0 otherwise.

\section{Numerical Results}
In this section, we evaluate the performance of the proposed semantic communication system and analyze the impact of the SKB on system performance. 
  
\subsection{Simulation settings}
The adopted dataset is the Caltech-UCSD Birds-200-2011 (CUB)\cite{cubdataset}, which contains 11788 images from 200 categories. There are 312 continuous attributes for each class. All attributes of all classes compose a 200$\times$312 matrix, which is exploited as the shared SKB in this paper. 9431 images are utilized for training and the remaining 2357 are assigned for testing. All images are resized as the dimension of 128 $\times$ 128 $\times$ 3.
Two benchmarks are considered as follows.
\begin{itemize}
    \item JPEG+LDPC: This is a traditional communication method, where JPEG and LDPC are adopted for source coding and channel coding, respectively. The generation task is substituted with reconstructing images and the classification task is performed based on the recovered images. 
    \item Vanilla SemCom: This is a vanilla semantic communication scheme with a shared training dataset, comprising an encoder, a decoder and a classifier subsequent to the decoder. The classification task is executed after image reconstruction. The network model is similar to the proposed scheme, where the encoder-decoder model is based on NVAE but with unconditional VAE, and the classifier is based on APN which outputs the class labels.
\end{itemize}

We evaluate our method and benchmarks in classification accuracy for image classification task, peak signal-to-noise ratio (PSNR), structure similarity index measure (SSIM), learned perceptual image patch similarity (LPIPS) and Fr\'{e}chet Inception distance score (FID) \cite{fid} for image generation task, and semantic accuracy for system performance. FID measures the distribution discrepancy between the reconstructed images and real images, which is extensively employed to assess the quality and diversity of generated samples. The SNR varies randomly from 0dB to 10dB in the training stage. 

\begin{figure}[t]
	\centering
	\includegraphics[width=0.6\linewidth]{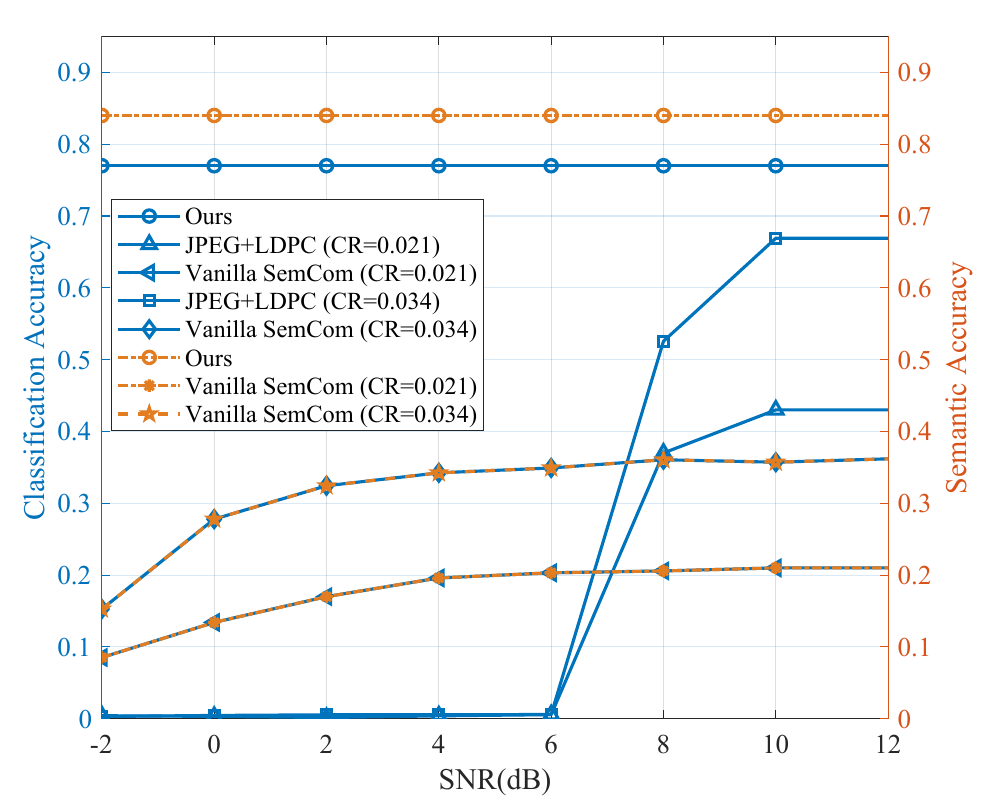}
	\caption{Classification accuracy and semantic accuracy versus SNR.}
	\label{classification task}
\end{figure}

\begin{figure}[t]
	\includegraphics[width=0.98\linewidth]{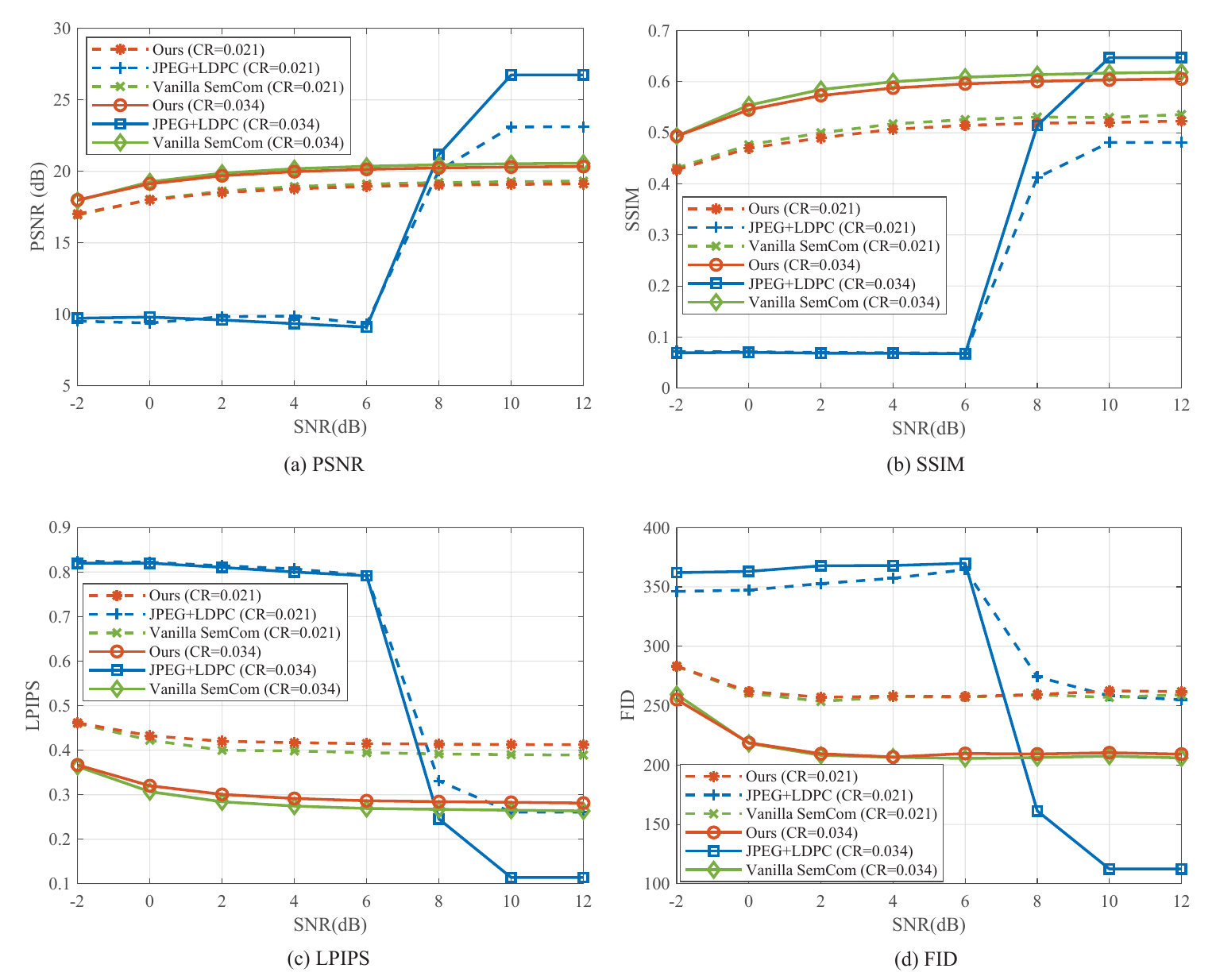}
	\caption{Performance comparison of (a) PSNR, (b) SSIM, (c) LPIPS and (d) FID within different SNR. Better performance is indicated by higher values of PSNR and SSIM, and lower values of LPIPS and FID.}
	\label{perfomance}
\end{figure}

\begin{figure}[t]
	\centering
	\includegraphics[width=0.98\linewidth]{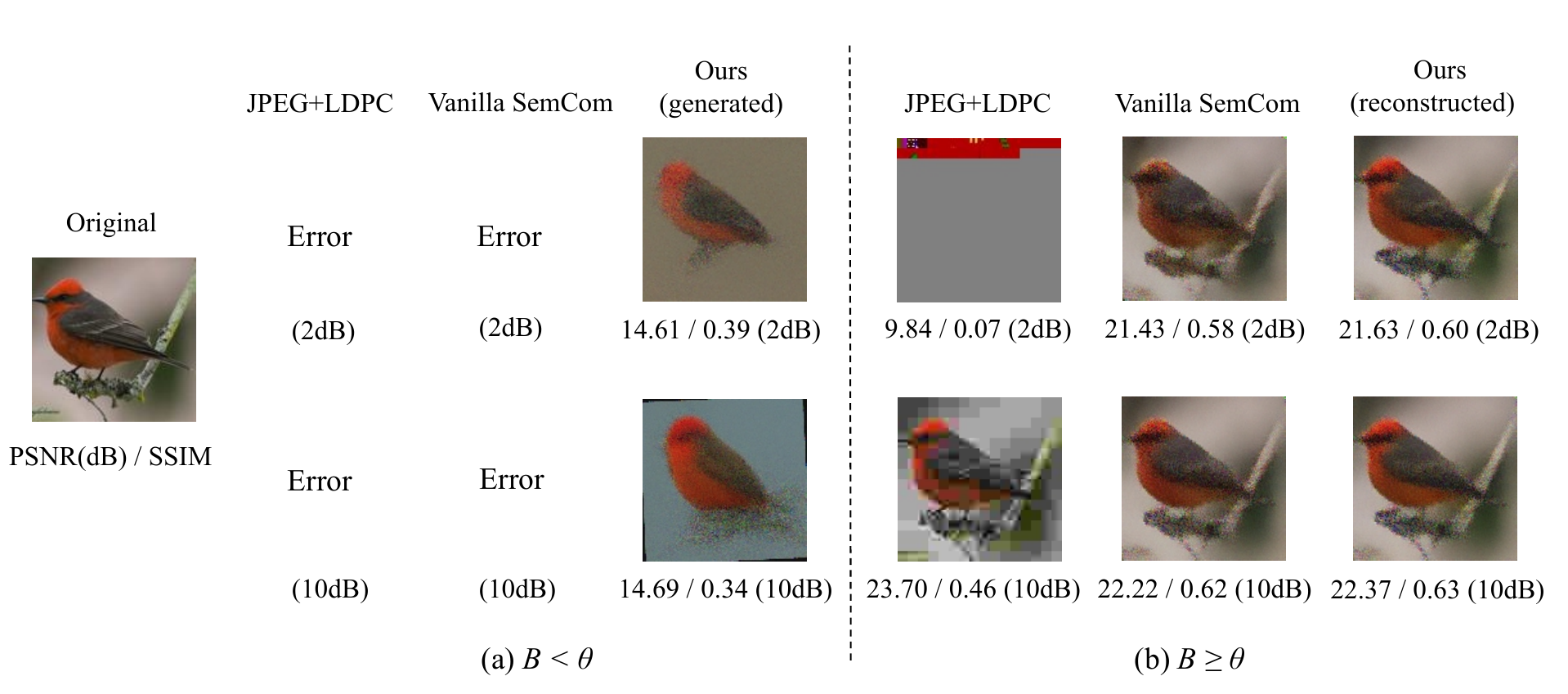}
	\caption{Examples of visual comparison when $\theta$ = 0.021. The first and second rows represent the cases with SNR as 2dB and 10dB, respectively. }
	\label{visualimg}
\end{figure}
\addtolength{\topmargin}{-0.13cm}
\subsection{Simulation results }
 Fig.~\ref{classification task} shows the classification accuracy and the semantic accuracy versus SNR when CR$\ =0.021$ ($B=\theta=0.021$) and CR$\ =0.034$ ($B=\theta=0.034$). Since the index $\mathbf{v}$ is transmitted without error, both accuracies of the proposed method are not affected by the SNR and CR, where the classification accuracy holds 0.77 and the semantic accuracy holds 0.84. The classification accuracy of the JPEG+LDPC method declines sharply when SNR is below 8dB, indicating the scheme fails to make inferences in the low SNR regime. The Vanilla SemCom has increasing classification accuracy with larger CR and SNR, while the precision is always lower than 0.4 and the gains obtained from higher SNR are minimal.

Besides, the semantic accuracy of our approach is obviously higher than classification accuracy, which validates that the receiver is able to obtain some useful information from erroneous outcomes, benefiting from the setting of SKB and semantic transmission. However, the Vanilla SemCom has equivalent classification accuracy and semantic accuracy. This is because the Vanilla SemCom solely relies on training with a shared dataset, and no valuable information can be acquired once the error occurs during the inference.

Fig.~\ref{perfomance} compares the PSNR, SSIM, LPIPS and FID of the proposed method with benchmark schemes against SNR. It can be observed that the proposed method outperforms the JPEG+LDPC scheme in the low SNR regime. Meanwhile, the proposed scheme has similar performance with the Vanilla SemCom, which demonstrates that with an efficient semantic encoder, the image reconstruction performance of CVAE-based model is on par with that of unconditional VAE-based model. Moreover, both deep learning-based methods only have slight performance degradation when SNR decreases, exhibiting resistance to the channel noise. 
It should be noted that Fig.~\ref{perfomance} depicts the image reconstruction performance of our method and shows the cases $B\geq\theta$ when $\theta=0.021$ and $0.034$, respectively. In the case of $B<\theta$, we use FID to evaluate the quality 
of generated images. The FID values with $\theta=0.021$ and 0.034 are 281.95 and 273.23 respectively, which are larger than those of image reconstruction but acceptable.

Some visual examples when $\theta=0.021$ are shown in Fig.~\ref{visualimg}. 
If the testing CR $B$ is below the training CR $\theta$, e.g., $B=0.001<\theta$ in Fig.~\ref{visualimg}(a), only indices are transmitted and our scheme yields recognizable images with same class attributes, while the JPEG+LDPC scheme is unable to compress the image with $B=0.001$ and the Vanilla SemCom is unsuccessful in transmitting only $0.001\times128\times128\times3\approx49$ symbols to recover the image.  When $B\geq\theta$, e.g., $B=0.021=\theta$ in Fig.~\ref{visualimg}(b), the proposed method provides favorable reconstructed images, which have better visual quality compared to the benchmarks. 

\begin{table}[ht]
\caption{Performance versus the size of SKB.}\label{skbsize}
\centering
\scalebox{0.95}{\begin{tabular}{c c c c c c c}
\toprule
$d$ &30 &78&156&234&312 \\
\midrule
Classification accuracy&0.37 & 0.75&0.77&0.77 & 0.77\\
Semantic accuracy&0.55 & 0.82&0.84&0.84 & 0.84 \\
PSNR(dB)                & 19.09& 19.03 &19.05  & 19.07  &19.09\\
SSIM                        & 0.52&0.53&0.53& 0.52 &0.52\\
LPIPS                    & 0.41&0.41&0.42	&0.41&0.41\\
FID                     & 259.90&256.90&258.35&257.82&262.39\\
\bottomrule
\end{tabular}}
\end{table}

In addition, the effect of the size of SKB is investigated and performance comparison when $B=\theta=0.021$ and SNR=10dB is given in Table \ref{skbsize}. The parameter $d$ represents the dimension of the attribute vectors in the SKB. It can be seen that the larger the SKB, the better the performance of the two tasks. Besides, when the size of SKB is reduced to 1/4 of the original size, the performance only experiences a slight decrease, which verifies the SKB still retains enough relevant information to support effective communications.

\section{Conclusions}
In this paper, we propose an SKB-enabled generative semantic communication system for image classification task and image generation task. The end-to-end semantic transmission process is powered by a concise SKB, which allows for the transmission of only the relevant indices. Benefiting from the guidance of the SKB, the image classification task is executed at the transmitter and controllable generation capability is equipped to perform the image generation task.
Besides, we utilize the accuracy of the transmitted semantics as a new metric to evaluate the system performance. Numerical results show that the proposed method exhibits superior performance with few transmitted load, especially in the low SNR regime.


\end{document}